# DNA origami

**Swarup Dey,[1] Chunhai Fan,[2,3] Kurt V. Gothelf,[4] Jiang Li,[2,5] Chenxiang Lin,[6] Longfei Liu,[7] Na Liu,[6] Minke A. D. Nijenhuis,[4] Barbara Saccà,[8] Friedrich C. Simmel,[9] Hao Yan,[1] Pengfei Zhan[10]**

1. Center for Molecular Design and Biomimetics, The Biodesign Institute, School of Molecular Sciences, Arizona State University, Tempe, AZ 85287, USA.
2. School of Chemistry and Chemical Engineering, Frontiers Science Center for Transformative Molecules and National Center for Translational Medicine, Shanghai Jiao Tong University, Shanghai 200240, China
3. Institute of Molecular Medicine, Shanghai Key Laboratory for Nucleic Acids Chemistry and Nanomedicine, Renji Hospital, School of Medicine, Shanghai Jiao Tong University, Shanghai 200127, China
4. Interdisciplinary Nanoscience Center (iNANO) and the Department of Chemistry, Aarhus University, 8000 Aarhus C, Denmark.
5. Bioimaging Center, Shanghai Synchrotron Radiation Facility, Zhangjiang Laboratory, Shanghai Advanced Research Institute, Chinese Academy of Sciences, Shanghai, 201204, China.
6. Department of Cell Biology and Nanobiology Institute, Yale University, West Haven, CT 06516, USA.
7. Max Planck Institute for Solid State Research, Heisenbergstrasse 1, 70569 Stuttgart, Germany.
8. Centre for Medical Biotechnology (ZMB), University of Duisburg-Essen, Universitätstr. 2, 45117 Essen, Germany.
9. Physics Department, Technische Universität München, 85748 Garching, Germany.
10. 2nd Physics Institute, University of Stuttgart, Pfaffenwaldring 57, 70569 Stuttgart, Germany.

*E-mail: fanchunhai@sjtu.edu.cn
The authors are listed in alphabet order.

## Abstract

Biological materials are self-assembled with near-atomic precision in living cells, whereas synthetic 3D structures generally lack such precision and controllability. Recently, DNA nanotechnology, especially DNA origami technology, has shown great utility in bottom-up fabrication of well-defined nanostructures ranging from tens of nanometres to sub-micrometres. In this review, we summarize the methodologies of DNA origami technology, including origami design, synthesis, functionalization and characterization. We highlight applications of origami structures in nanofabrication, nanophotonics/nanoelectronics, catalysis, computation, molecular machines, bioimaging, drug delivery and biophysics. We identify challenges for the field, including size limits, stability issues, and the scale of production, and discuss their possible solutions. We further provide an outlook on next-generation DNA origami techniques that will allow in vivo synthesis, and multiscale manufacturing.

# 1. Introduction

Biological materials are synthesized in a bottom-up manner in living cells, which exploits the information encoded in biomolecules to guide their self-assembly to form hierarchical structural complexes spanning from nanometres to the macroscopic scale with near-atomic precision. By contrast, in-vitro manufacturing of complex 3D structures down to the nanometre scale until now lacks such precision and controllability.

In the 1980s, Seeman first demonstrated the rational design of an immobile Holliday (4-arm) junction[1], which turned DNA into a nanoscale polymer extending in two dimensions, beyond the capabilities of simple one-dimensional (1D) double helices, which was regarded as the debut of the field of DNA nanotechnology. Since then, various types of DNA nanostructures have been fabricated, including double-crossover (DX) DNA tiles[2], triple-crossover (TX) tiles[3], $4 \times 4$ tiles[4], and three-point-star tiles[5]. Acting as supramolecular building blocks these tiles can be further assembled into higher order nanostructures such as 2D lattices, nanotubes, and more complicated 3D structures such as polyhedra, hydrogels and crystals[6, 7]. As a bottom-up self-assembly approach, DNA nanotechnology allows massively parallel synthesis of well-defined nanostructures, e.g., a picomole-scale synthesis can generate $10^{12}$ copies of a product.

DNA origami technology, as a promising branch of structural DNA technology, has shown great utility in bottom-up fabrication of well-defined nanostructures ranging from tens of nanometres to sub-micrometres. The concept of DNA origami first invented by Rothemund in 2006[8] relies on folding a long single-stranded (ss-) DNA (scaffold, typically a M13 bacteriophage DNA of ~7,000 nucleotides long) with hundreds of designed short ssDNAs (staples) via Watson-Crick base pairing. Compared to tile-based DNA assembly strategies, DNA origami synthesis exhibits higher yield, robustness and more power to build complex non-periodic shapes. Since the original demonstration of two-dimensional (2D) patterns[8], almost arbitrary shapes have been synthesized, from one-dimensional (1D) to three-dimensional (3D) structures with user-defined asymmetry, cavities, or curvatures[7, 9]. More recent progress includes the hierarchical assembly of supersized structures[10, 11], single-stranded origamis[12, 13], and dynamic structures[7, 14, 15] (Fig. 1). Further, the programmability of DNA origami allows computer-aided design and universal synthesis protocols[16], making DNA origami an easy-to-use technology, amenable to automated fabrication.

Typical planar DNA origami structures contain approximately 200 uniquely addressable points in an area of 8,000–10,000 square nanometers[7]. This global addressability with nanometre resolution allows them to serve as elaborate templates or frameworks, which can precisely localize multiple desired functional moieties and materials (e.g., fluorescent dyes, metals, minerals, polymers, carbon nanotubes, functional nucleic acids, enzymes, therapeutic molecules) in well-defined numbers and patterns[6, 7, 17], which have shown great promise in the fabrication of structures enhanced by metal, silica, lipid or polymer coatings[18-20], and as nanosystems for super-resolution fluorescence microscopy, light harvesting applications, photonic transfer, plasmonic chirality, surface-enhanced Raman spectroscopy (SERS), and nanoelectronic circuits[21-26]. Moreover, dynamic DNA origami structures can be rationally engineered on the basis of reconfigurable modules, which utilize strand displacement reactions[14], conformationally switchable domains and base stacking components[15], enabling a variety of applications such as target-responsive biosensing and bioimaging[27], smart drug delivery[28, 29], biomolecular computing[30] and nanodevices allowing external manipulation with light or other electromagnetic fields[31-33].

In this review, we summarize the methodologies of DNA origami technology, including origami design, synthesis, functionalization and characterization. We highlight applications of origami structures in nanofabrication, nanophotonics/nanoelectronics, catalysis, computation, molecular machines, bioimaging, drug delivery and biophysics. We identify challenges for the field, including size limits, stability issues, and the scale of production, and discuss their possible solutions. We further provide an outlook on next-generation DNA origami techniques that will allow in vivo synthesis, and multiscale manufacturing.

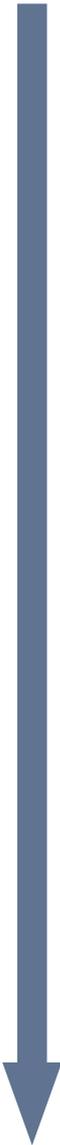

**Figure 1. A timeline of representative advances in the field of DNA origami.**

## 2. Experimentation

DNA origami objects with rich diversity in dimension, geometry, and shape have been developed, ranging from single layers to multilayers[8, 34, 64] as well as from flexible wireframes to rigid polyhedroa[40, 46, 65]. A typical experimental process of fabricating DNA origami is illustrated in Figure 2.

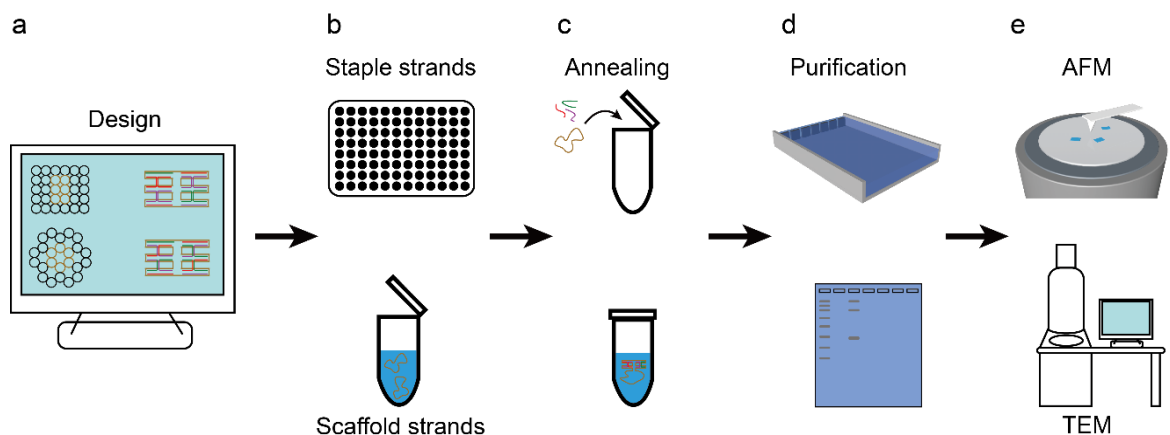

**Figure 2. General principle of the DNA origami design and assembly.** (**a**) DNA origami structures are usually designed using caDNAno software. (**b**) Staple strands are usually purchased from commercial and stored in 96-well plates. Single-stranded M13 DNA is typically used as a scaffold for DNA origami structures. (**c**) M13 scaffold mixed with staple strands (with a large excess) is assembled through thermal annealing. (**d**) The structures are usually purified using agarose gel electrophoresis. (**e**) Atomic force microscopy (AFM) is usually used to observe two-dimensional and single-layer origami structures. Transmission electron microscopy (TEM) is often used to characterize three-dimensional origami structures and multilayer structures.

**Step 1: Design of DNA origami**

The inventions of DNA origami benefitted from the development of computer programs[8], which gradually evolved into a family of softwares for the design and analysis of complex DNA origami nanostructures.

Table 1 presents a comparative summary of the different software developed for designing DNA origami structures. The first generation (1G) DNA origami design softwares (Table 1, grey highlighted rows) were developed for designing various 2D and 3D origami structures. Detailed insights on designing origami by 1G software have been covered extensively by Dietz *et al.*[66], and can also be found in the references cited in Table 1. caDNAno remains the most reliable software for designing DNA origami. Other software such as Tiamat, SARSE-DNA, Nanoengineer-1, Hex-tiles, GIDEON, K-router, etc. have also been used for DNA origami designs. These platforms are mostly used by experts in the field.

**Table 1. Comparison between different DNA origami design software.**

| Software | Ability | Advantages | Disadvantages |
|---|---|---|---|
| **a) caDNAno[67]** | Lattice-based (honeycomb or square) scaffolded DNA origami design. | • **Simplest user interface among all 1G design software.**<br>• **Semi-manual inter-helix crossover creation.**<br>• **Most widely used. Strong community and developer support.** | • **Difficult to design 3D structures due to the lack of a 3D interface with a single-base resolution.**<br>• Difficulty of downstream design modification, e.g. single-stranded staple overhangs.<br>• **Not suitable for non-parallel helix based-structures, e.g. - 3D wireframes, single-stranded tiles etc.** |
| **b) Tiamat[68]** | DNA nanostructure design without lattice or scaffold limitations. | • **Versatile, hence more suitable for almost any structures, ranging from wireframes, single-stranded tiles, DNA crystal motifs, etc.**<br>• Parallel, non-parallel and branched helix designs.<br>• 3D workspace with single-base resolution.<br>• Easy downstream design modifications.<br>• Custom sequence generator with options to vary G-C%, unique sequence limit, G-repetition, etc. which can be used for other applications as well. | • **Manual creations of crossovers.**<br>• **More expert knowledge is required.**<br>• **Less widely applied than caDNAno, when simple designs are needed.**<br>• Only windows version but no mac OS version available.<br>• Weak community support. |
| **c) SARSE-DNA[69]** | Lattice-based DNA origami design. | • Very similar to caDNAno.<br>• Option to export as all atom PDB format for MD simulations. | • No parallel helix based-structures such as wireframes, single stranded tiles, etc.<br>• **Very similar to caDNAno for scaffolded origami designs.** |
| **c) Nanoengineer-1** | DNA nanostructure design without lattice or scaffold limitations. | • Very similar to Tiamat.<br>• Option to export as all atom PDB format for MD simulations. | • Crossover creation is manual similar to that for Tiamat.<br>• **Scarcely used, and hence weak community support** |
| **d) Hex-Tiles[70]** | Triangulated wireframe SST structures | • SST-wireframe design.<br>• Custom arrangement of sequences in 96 well plates according to experimental convenience.<br>• Re-usage of existing strands. | • Limited to SST structures.<br>• Not used widely in the community. |
| **e) v-Helix[46] (*)** | 2G automated design of scaffolded 3D DNA origami with single-dual duplex edges. | • **Automated design of complex 3D structures from user-drawn polygonal meshes.**<br>• No parallel helix required.<br>• **Only design platform with in-built relaxation algorithms to predict the folding of complex polyhedral structures.**<br>• The minimum use of DNA is ensured. Cost-effective and low salt stable structures. | • **Polygonal wireframe structures must be topologically equivalent to a sphere. Hence, not suitable for simpler 1D, 2D structures or more complex 3D structures.**<br>• Rely on Autodesk Maya or other 3D design software to create the mesh design. Different interface requirements for design, relaxation and sequence generation.<br>• Relatively new, less widely tested across labs compared to caDNAno and Tiamat. |
| **f) DAEDALUS[16]** | 2G automated design for scaffolded 3D DNA origami with dual duplex edges. | • **Automated design without limitation to spherical topologies**.<br>• Less material consuming wireframe designs ensure cost-effective and low salt stable structures. | • **No integrated relaxation algorithms to predict structure folding.**<br>• Each arm must contain at least two helices, and hence more material intensive than v-Helix. |
| **g) TALOS[71]** | 2G automated design for scaffolded 3D DNA origami with honeycomb six-helix edges. | • **Automated design of mechanically stiff structures.**<br>• Control over distribution of staple lengths. | • Material intensive and incompatible to low-salt, physiological buffers. |

The second generation (2G) design softwares are more advanced and can generate sequences in an automated fashion from user-provided 3D designs (Table 1, blue highlighted rows). vHelix[46] is the most widely used software in this category. It also contains an integrated simulation platform that can predict the folding of the designed structures in solution. Other software includes DAEDALUS[16], TALOS[71], etc. Despite being mainly developed for complex 3D structures, the automated version for 2D designs was also released later[72, 73]. Other automated design software such as MagicDNA (Castro et al., in preparation), Adenita[74] is still under development. Such software is mainly developed for non-experts. Hence, it is more user-friendly and demands less technical knowledge than their 1G counterpart. However, it is more recent, and hence less widely tested by the community.

DNA origami creation is costly and time-consuming. Hence, it is important to predict the folding of the designed origami computationally. Due to the substantial sizes of DNA origami structures and the timescales of complex events, such as hybridization and dehybridization, which take place at micro to millisecond scales, it is computationally too challenging to use conventional all-atom MD simulations for DNA origami structure predictions. Various simulation software packages have been developed to predict the folding of DNA origami structures after designing. OxDNA[75-77] offers the most versatile and practical approach in terms of ease of usage and features, whereas the very recently developed platform MrDNA[78] offers the highest resolution as well as the fastest speed. The bottleneck, however, is requirement of CUDA enabled GPU (in a supercomputing cluster) to run simulation which is quite resource intensive and needs significant coding knowledge. The choice of appropriate simulation packages is often determined based on the target applications and availability of resources. Completely web-server based coarse graining packages such as CanDo[79] and COSM[80] can also be used to predict the folding or mechanical strain in the designed origami structure.

**Step 2: Assembly of DNA origami**

A long single-stranded circular viral genome is used as a "scaffold". The choice of the scaffold is determined by the size and complexity of the desired structure. The most commonly used scaffold is the 7249nt long m13mp18 viral genome. Other typical scaffolds include p7308, p7560 and p8064. They can be purchased from companies, such as New England Biolabs, Guild Biosciences, Tilibit Nanosystems, etc. or custom made using asymmetric PCR[81], enzymatic single-strand digestion of PCR-amplified double-stranded DNA[34] or by purifying phage-derived single-stranded genomic DNA[34].

After exporting from the design software as .csv or text files, the short single-stranded "staple" strands are mainly purchased in 96 well plate formats from Integrated DNA Technologies. Ordering as 96-well plates allow the staples to be pipetted using 12 or 8-channel pipettes (Eppendorf) for the mixture (Figure 1b). Normally, the concentration of the purchased staple strands is about 100-150 µM and they should be stored at -20°C in 5 mM Tris base and 1 mM EDTA at pH 8.

Most protocols for the assembly of DNA origami involve pH=8 Tris-acetate-EDTA (TAE) buffers with different concentrations (5 mM to 20 mM) of $Mg^{2+}$ ($MgCl_2$). The yield of DNA origami is sensitive to the concentration of $Mg^{2+}$. The exact concentration of $Mg^{2+}$ may depend on the complexity of the DNA origami. The most commonly used buffer contains 12.5 mM $Mg^{2+}$. Higher concentrations (16.5-20 mM) are used for 3D structures, whereas lower concentrations (5-10 mM) are used for wireframe origami or tiles. It should be noticed that DNA origami assembled with high $Mg^{2+}$ concentrations can be transferred into buffers with low $Mg^{2+}$ concentrations, while keeping the structural intact.

DNA origami can be folded via one-pot self-assembly. As shown in Figure 1c, the folding mixture contains scaffold DNA, staple strands, pH stabilizing buffer with cationic and ultra-pure water. Table 2 presents the best practices for working with the reagents necessary to create DNA origami. In general, to reduce non-specific aggregates, the concentration of the scaffold strands ranges from 5-20 nM, and the concentration of the staple strands is about 10 to 20 times higher than the scaffold strands. For dynamic DNA structures, the staple to scaffold ratio should be optimized for staples depending on the dynamic functions. Generally, it is 1.5-2 times. After mixing the components, generally, a 10-100 µL mixed solution undergoes a thermal annealing process in a PCR thermocycler. Typically, the annealing process is done in 2 steps – 1) heating the mixture to boiling temperature (85 or 90 °C) shortly (5-10 mins) to disrupt all secondary structures, 2) gradual cooling from 85 °C to 25 °C for the strands to spontaneously self-assemble to form DNA origami. The specific annealing procedure depends on the complexity of the DNA origami – small wireframe structures and 2D origami need a few hours, while multilayer 3D structures may require several days.

**Table 2. Best practices for working with the reagents necessary to create DNA origami.**

| Component | Best practices |
|---|---|
| Scaffold | • While designing an origami, consume as much scaffold as possible. Keep the unused scaffold part no bigger than 100-200nts at one end of the desired structure.<br>• Keep scaffold frozen at -20 or -80ºC as small aliquots. Avoid frequent freeze-thaws. |
| Staples | • Order in 96 well plates.<br>• Store plates at 4 ºC for a short term, -20 or -80ºC for long-term storage. Avoid frequent freeze-thaws.<br>• Purify important staples by denaturing PAGE gel or HPLC. This ensures perfect incorporation of the strands into the origami.<br>• Order modified staples separately from normal unmodified staples. This makes it convenient to anneal the structures.<br>• Staple mixes can be created for different parts of the origami for frequent annealing. This avoids the need to repeatedly freeze-thaw the source plates. |
| Folding buffer | • Use freshly prepared buffers.<br>• For buffers to be used for sensitive microscopic techniques such as AFM or TEM, filter the buffers using syringe filters and store them in glass vials. This avoids leaching of plastic fibers. |

## Step 3: Purification of DNA origami

Purification and enrichment are crucial steps to use DNA origami for biochemical and biomedical applications. A detailed overview of the purification methods has been covered elsewhere by Dietz *et al*.[66] and is beyond the scope of this primer. For various applications of DNA origami, the optimal purification methods should be chosen by comparison of five methods based on two quantitative (yield, duration) and four qualitative (volume limitation, dilution, residuals, damage).[82] For example, PEG precipitation are adapted to enrich target species with high yield, but it also introduces residual PEG molecules; Filter purification with molecular weight cut-off membranes provides residual-free separation but with volume limitation; Gel purification is suitable for band-pass molecular weight separation, as the complex of DNA origami and nanoparticles.

## Step 4: structural analysis of DNA origami

The DNA origami structures can be characterized using microscopic techniques, such as gel electrophoresis, transmission electron microscopy (TEM), scanning electron microscopy (SEM) and atomic force microscopy (AFM) (Figure 1e). It should be noted that AFM may not be suitable for the imaging of 3D or multilayer DNA origami because deformation caused by AFM tips makes it difficult to obtain a complete topography when scanning a surface feature, especially for low rigid structure, but it is powerful to characterize single-layer or many 2D DNA origami structures. With the help of uranyl formate (Science Services), negative-stain TEM is a useful tool for the structural characterization of 3D or multilayer DNA origami. The image processing (EMAN2 software) could be used to assess the heterogeneity of DNA origami, enabling the identification of the structural flaws and reconstruction of 3D models from TEM images of a single structure (https://blake.bcm.edu/emanwiki/EMAN2). Table 3 shows a comparative summary of the different structural analysis methods for DNA origami structures. Additionally, Föster Resonance Energy Transfer (FRET) is also used to analyse dynamically reconfigurable structures[83, 84].

**Table 3. Comparison of different structural analysis methods and best practices.**

| Method | Advantages | Disadvantages | Best practices |
|---|---|---|---|
| Gel electrophoresis | • Much simpler and resource friendly compared to microscopic methods.<br>• Bulk estimation about structure formation yield, purity, structural reconfiguration etc. is possible.<br>• Additional option of high-resolution purification of the desired population. | • Cannot provide in-depth structural insights such as microscopic techniques. | • Buffer used to cast gel, running buffer and sample buffer should ideally be same.<br>• Gel overheating must be avoided to prevent gel melting and sample degradation. Always run at cold condition ~ 0-4 ºC. Avoid monovalent ions > 100mM in running buffer even if sample buffer has that.<br>• Intercalating dyes tend to fall off after 2-3 hours, making it difficult to visualize the structure. Either use smaller gel % or use fluorescently labelled strands if it is unavoidable to run > 3hours.<br>• Additionally, just for structural analysis, especially for fluorophore conjugated samples, addition of intercalating dyes should be avoided for high resolution analysis. For example – dynamic structures that have FRET in one configuration can be differentiated by intensity of donor fluorophore in this mode. |
| Transmission electron microscopy | • Best for 3D structures.<br>• Highest resolution.<br>• Can be coupled with particle averaging methods to enhance resolution.<br>• Especially suitable for metal nanoparticle modified structures | • Negative-staining is cumbersome and time consuming<br>• Difficult to obtain the sample height information.<br>• Structural deformation due to drying. | • TEM grid should be discharged first to enhance the hydrophilicity.<br>• Before staining, add NaOH solution to adjust the pH of the uranyl formate solution.<br>• EMAN2 software could be used to assess the heterogeneity of DNA origami. |
| Atomic force microscopy | • **High fidelity**<br>• Check in solution or air<br>• Modified AFM tip can be used to study effect mechanical force on DNA.<br>• High speed AFM can probe structural dynamics in real time. | • Unsuitable for 3D or multilayer DNA origami.<br>• Time-consuming. | • Depositing nickel acetate first to prevent the movement of sample. The amount of $Ni^{2+}$ should be carefully adjusted for origami type – smaller structures need higher $[Ni^{2+}]$ to be immobilized on surfaces.<br>• Purified samples give best results.<br>• Using deionized water to get rid of salts when performing in air.<br>• Tapping mode is more useful. SNL-10 tip is recommended.<br>• Peak-force amplitude should be adjusted depending on the structural flexibility. Typically used values are 20-30nm for fluid mode and 100-150nm for air mode. |
| Scanning electron microscopy[85] | • Fast<br>• No need of negative-staining<br>• Accompanied by analysis | • Low resolution<br>• Unsuitable for 3D or multilayer DNA origami. | • Samples could be prepared on mica or silicon wafers.<br>• Use deionized water to get rid of salts, and keep dry.<br>• Use a low voltage mode. |

# 3. Results

In the previous section, we have briefly discussed the experimental route from the design to the assembly and purification of DNA origami structures. Here we provide typical characterization data for the structure and function of DNA origami assemblies with the ensemble and single-molecule techniques (Fig. 3). Efforts in this direction have led to a constant improvement in the quality and quantity of the information attainable from the self-assembly process, enabling to test hypotheses, refine procedures and generate new ideas in the strive for optimization. In this sense, the critical analysis of the results of the self-assembly reaction has probably represented the main boost that drove the DNA origami nanotechnology to the current state-of-the-art.

**Ensemble Characterization.** To this class belong methods such as agarose gel electrophoresis (AGE) mobility assay, UV-Vis and fluorescence spectroscopy and circular dichroism. Here, the outcome of the assembly reaction is examined as a whole and information is obtained on the relative amount of the various components of the solution mixture. AGE is undoubtedly the "workhorse" of all ensemble techniques and the first analysis to be performed to prove whether the desired structure is indeed present and to which extent[34, 86-88]. Upon application of an electric field, DNA molecules migrate along a polymer gel matrix according to their size, charge and shape, enabling to separate multimers of different orders, as well as misfolded and/or partially assembled intermediates (Fig. 3a). DNA is then visualized staining the gel with an intercalant dye that fluoresces under UV-light illumination and products are quantified using fluorescence gel-scanners and modern software tools. The attachment of fluorescent dyes at defined locations of the structure may be advantageously used to uniquely identify and isolate the product of interest[89]. In particular, when photoactive compounds capable of FRET are positioned at defined intermolecular distances within the DNA structure, dynamic processes can be revealed in real-time (Fig. 3b-d) [36, 90, 91, 92, 93, 94]. The effective actuation of various DNA devices has been proven and the kinetics of the induced transformation has been fully characterized with FRET[15]. Ensemble techniques are therefore valuable tools to gather a global overview of the process; their major limitation however is a lack of sufficient spatial-temporal detail that might be indeed relevant at the microscopic scale.

**Single-Molecule Characterization.** The main feature of a DNA origami structure is to provide a molecular surface where desired chemical species can be positioned at predictable nanometre distance with an accuracy of about 5 nm. Clearly, the best way to prove and exploit this peculiar structural property is through single-molecule techniques. Both force-and optical-based methods have been employed to characterize the structure and function of DNA origami objects[95-97], namely, AFM, TEM, single-molecule (SM) fluorescence microscopy and more recently, SM-force measurements. The very beginning of the DNA origami era has been undoubtedly portrayed by eye-catching AFM images that clearly demonstrated the success of the method and spurred other laboratories to explore the ground (Fig. 3e)[8]. By sensing the intermolecular forces occurring between the tip and the sample, AFM provides the height profile of the specimen deposited on an atomically flat surface, i.e. its detailed topographical map, with a lateral resolution down to 1-2 nm. The capability of this technique to reveal fine structural features with sufficient detail has been recently employed to shed more light into the folding pathway of planar DNA origami structures[98-101]. Such high-spatial resolution is nevertheless maintained till the micrometre scale, providing accurate wide-field images of hierarchically self-assembled constructions (Fig. 3f)[39, 102]. Decoration of the DNA origami surface with bulky molecules, such as biotin/streptavidin markers, hairpin dumbbell motifs or nanoparticles, attached to pre-selected positions of the architecture was initially applied to prove the potential use of such structures as molecular pegboards.[8] Later, this strategy has been applied to break the symmetry of regular origami shapes, such that their relative orientation in respect to the mica support, as well as the occurrence of molecular events at different locations of the same structure, could be unambiguously determined (Fig. 3g).[37, 38, 103-105] Finally, modern AFM instrumentations allow nowadays to combine a high spatial resolution with a temporal resolution of seconds to subseconds[95], which is sufficient to monitor the change in the topology of single molecules (Fig. 3h)[106] or the action of DNA processing enzymes in real time.[107]

Whereas AFM is more suitable for mono- and two-dimensional structures, TEM and cryo-EM are instead preferentially used to characterize three-dimensional DNA objects. First TEM images of 3D DNA origami structures showed the suitability of this technique to reveal the successful formation of the intended space-filled architectures (Fig. 3i)[34, 35]. However, the high vacuum and dehydration conditions associated to TEM imaging may result into flattening and distortion of structures that display inner cavities. In such instances, EM imaging in fully hydrated cryogenic conditions is preferred, as it ensures structure preservation and enables observation of the macromolecule in the close-to-native state in solution. Using cryo-EM, the first pseudo-atomic model of a 3D DNA object was obtained with an overall resolution of 11.5 Å, confirming the formation of the structure with the expected topology[108]. Although the information level contained in raw EM images is normally relatively low, the acquisition of large sets of individual particle images and their semi-automatic processing through sophisticated post-imaging software tools results in a dramatic increase of the signal-to-noise ratio and in most cases ends up in the full reconstruction of the 3D structure with almost atomic resolution. Similarly to AFM, TEM provides a wide-view image of the probe and has been therefore successfully employed to characterize the formation of micrometre-large DNA origami assemblies obtained by base stacking, guided hybridization or a combination of both[10, 55]. Finally, although being limited by a lower time-resolution, TEM can be still advantageously used to discriminate structurally distinct states, that become more populated upon environmental changes in pH, salt concentration or temperature, thus allowing to decipher the triggered dynamics of complicated molecular machines[15].

Fluorescence-based techniques provide an indirect characterization of local molecular events occurring nearby the dye molecules. These are attached to the DNA surface using various bioconjugation methods. As the DNA objects are typically smaller than the diffraction limit of light (ca. 200 nm), their resolution by optical means is possible only when the fluorescence signal has single-molecule sensitivity. Examples of these techniques are single-molecule FRET microscopy and super-resolution microscopy. Whereas FRET relies on the distance-dependent energy transfer between a donor and an acceptor photo-active probe, the basic principle behind super-resolution imaging is the consecutive switching of fluorescent molecules between an ON and OFF state. Stochastic reconstruction methods, such as point accumulation for imaging in nanoscale topography (PAINT)[97] have been smartly combined with the transient binding of short fluorescent DNA strands (DNA-PAINT) for the direct observation of dynamic events on DNA origami scaffolds (Fig. 3j)[45, 109]. In general, SM-fluorescence techniques have been successfully implemented to describe local dynamic events and quantify distance-dependent molecular interactions, conformational dynamics and kinetics of diffusion processes (Fig. 3k) [56, 104, 110, 111].

Of still limited use, but promising exciting applications in the field, is single-molecule force spectroscopy based on optical tweezers. In a typical optical trapping experiment, a few micrometer-large polystyrene sphere is trapped in the focus of a laser beam. Every small excursion from the focus leads to an imbalance of forces that drives the particle back to the equilibrium position. Such restoring force and its linear range-of-action are essentially the quantities sensed by the optical tweezer instrument enabling to monitor structural transitions of individual molecules with sub-pN resolution along nanometer distances. This technically challenging method has been already employed to observe the unzipping of small DNA origami domains[100, 112, 113] or the kinetics of G4 unfolding within a DNA origami cavity[114] and, particularly when in combination with optical methods[115], promises to be an essential tool for the investigation of dynamic events at the single-molecule level (Fig. 3l).

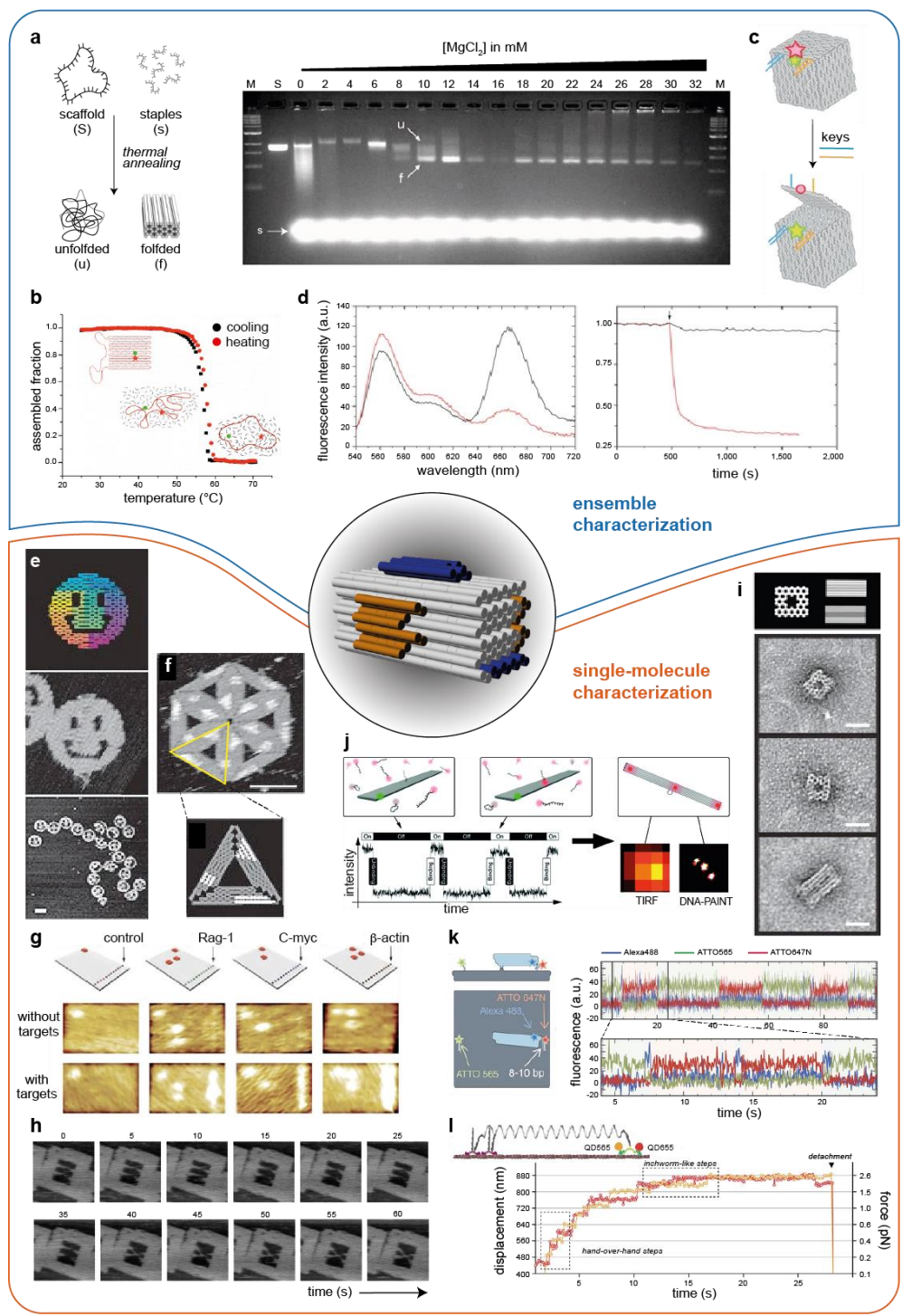

**Figure 3. Representative examples of ensemble and single-molecule characterizations of DNA origami structures.** (**a**) Agarose gel electrophoresis of the self-assembly products obtained at increasing magnesium ion concentrations[88] (from 0 to 32 mM). Undesired large aggregates are visible as non-migrating bands trapped into the loading pocket, whereas multimers of different orders typically result in the appearance of smears or slow-migrating defined bands (lanes 18-32). The formation of misfolded and/or partially assembled

intermediates (indicated as "u") can be inferred by the appearance of bands that migrate faster than the scaffold (lane S) but normally slower than the fully folded target structure (indicated as "f"). The excess of staple strands non-integrated within the origami structure are instead the faster migrating species recognisable at the bottom of the gel (indicated as "s"). (**b**) Thermal-dependent FRET spectroscopy has been used to monitor the thermally-driven assembly and disassembly of DNA origami microdomains[94]. Application of model-dependent methods allows to extrapolate the thermodynamic parameters of the process, gaining information on the structural features of the macromolecule. (**c-d**) Isothermal FRET spectroscopy coupled to single-strand displacement reactions may reveal the dynamics of local processes, which in turn can be employed as the signature of more global structural reconfigurations[36]. (**e**) First AFM images of quasi-planar DNA origami structures strikingly demonstrated the huge potential of the DNA as a construction material[8]. Scale bar is 100 nm. (**f**) Decoration of the DNA origami surface with bulky molecules leads to a change in the height profile of the structure at predictable positions[8]. This allows to visualize the formation of programmed hierarchical assemblies (scale bar is 100 nm) and (**g**) the occurrence of local binding events at predicted positions[105]. (**h**) The topological reconfiguration of G-quadruplex motifs (as well as other dynamic events in the second to sub-second regime) can be temporally resolved at the single-molecule level using fast-scanning AFM[106]. (**i**) First EM characterization of a space-filled 3D DNA origami structure[34] (raw images are provided in which the DNA object is visualized under different perspectives; scale bars are 20 nm). (**j**) Scheme showing a DNA origami structure with attached reference dye (green) and imager strands (red) in solution[45]. The docking strand extension is shown at the center of the structure. Without binding of the imager strand, no fluorescence in total internal reflection (TIR)-mode is observed (fluorescent OFF-state). Upon hybridization of an imager strand to the docking strand on the DNA origami, a bright fluorescence in the red channel is observed (ON-state). The positions of the fluorophores within a diffraction-limited TIRF area are then histogramed to obtain the reconstructed super-resolved DNA-PAINT image. (**k**) Single-molecule multicolor FRET experiments enabled to visualize the stochastic switching of a rotating DNA origami arm between two docking sites[56]. Fluorescence traces were obtained from three fluorophores: a donor (blue) attached to the six-helix bundle arm and two acceptor fluorophores (green and red) on opposite sides of the plate. The change between green and red fluorescence indicates switching of the arm between corresponding docking sites. (**l**) In a SM-force setup, a 200 nm fluorescent bead tagged with a DNA origami nanospring is optically captured and the cover glass is then moved by a piezo actuator[115]. Myosin VI is tethered to the nanospring and its unidirectional movement along the actin track fixed on the glass surface applies a force against the load of the nanospring. Coupling the SM force-signal to the optical readout derived by two different colours QDs attached to both heads of the protein revealed the stepping behavior of the motor, with both hand-over-hand and inchworm-like steps (dashed-line box).

# 4. Applications

DNA origami enables fast prototyping and precise engineering of molecular geometry, mechanics, and dynamics. With proper chemical modification at specific locations, DNA-origami structures provide a versatile engineering platform where nanoscale entities — from small-molecule dyes to massive protein complexes, from inorganic nanowires to three-dimensional liposomes — can be manipulated in highly programmable manners. In principle, engineering of any system where the collective outcome depends on molecular organization can benefit from the DNA origami-based engineering platform. Prominent examples include nanomaterial fabrication that relies on the exquisite control of molecular placement, as well as the study of biological processes resulting from well-organized molecular assemblies. In this section, we review selected DNA-origami-based applications in material science, physics, engineering, and biology, with the hope that these developments in the last decade could offer a sneak peek into the technology's broad impact in years to come.

**Nanofabrication**

Owing to their highly customizable geometric properties (e.g., size and shape) and nanometre resolution addressability, DNA origami structures have been used as templates or frameworks for the assembly or synthesis of diverse materials with nanometre precision, and have thus shown great promise in the nanofabrication of inorganic (metallic or non-metallic), polymeric, and biomolecular assemblies and patterns, with enhanced structural stability and/or desired physicochemical properties. Importantly, the DNA origami-based approaches allow massive parallel fabrication (e.g, $10^{12}$ copies of products in a single operation)[7] either in solution or on surface.

A representative nanofabrication approach (Fig. 4a) employs DNA origami templates with site-specific anchors to attach other prefabricated nanomaterials via user-defined, specific interactions. This approach enables the highly programmable arrangement of discrete nanostructures with up to atomic resolution, generating nanoassemblies in solution (colloids) or on surface (patterns), including metal and semiconductor nanoparticle assemblies with prescribed heterogeneity, anisotropy and/or chirality[41, 116-119], carbon nanotubes (CNTs) with defined alignments[22, 120, 121]. Their further packing or assembly generates higher ordered two-dimensional patterns[122, 123] or three-dimensional superlattices[48, 124]. Moreover, due to the programmable reconfigurability of DNA origami, the nanoassemblies can be dynamically rearranged, allowing responsively tuneable physicochemical properties[21, 118, 125].

In the in-situ synthesis approach (Fig. 4b), precursors (e.g., metal ions, silification precursors, lipid molecules) in solution adsorb/react/deposit on DNA origami templates with or without prescribed nucleation sites[126-128 59, 129], generating continuous architectures shaped by the morphologies of the templates or their cavities[44, 130]. This approach allows nanoarchitectures with almost arbitrary, user-defined geometries in solution or on surface[18, 59, 129, 131].

DNA origami structures and their derivatives have also been employed as masks or stamps for nanolithography (Fig. 4c), enabling the transfer of prescribed 2D nanoscale patterns onto other 2D or 3D materials [132, 133 134 23] with high fidelity. In this way, the size and shape of 2D nanomaterials (e.g., graphene) can be precisely tailored, allowing for the fabrication of diverse electronic devices with nanometre resolution. In addition, the combination of traditional lithography and DNA origami-enabled site-specific assembly has facilitated large-area spatially ordered arrays of functional nanoparticles and molecules on surface[135-137].

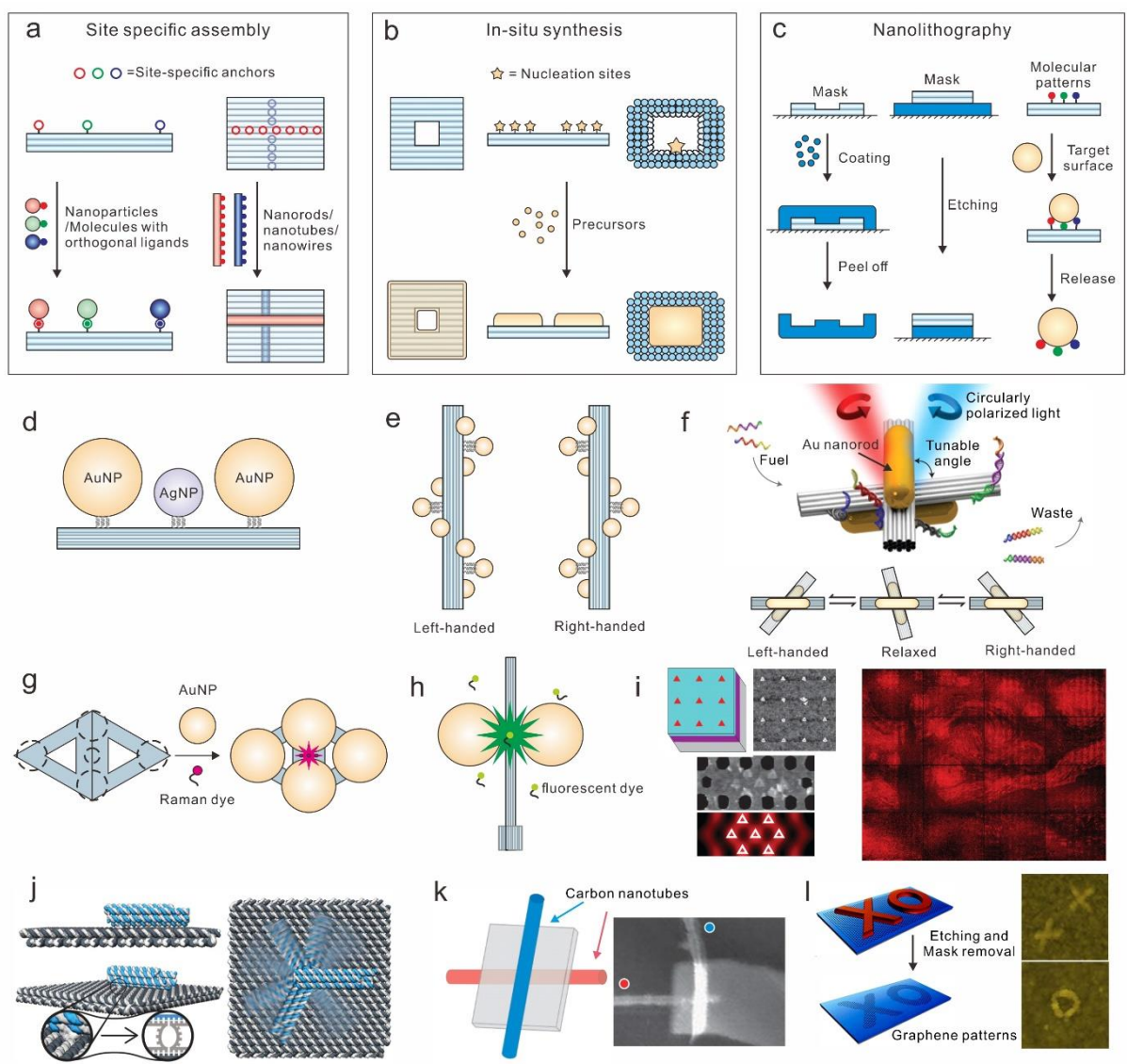

**Figure 4. Typical approaches of DNA origami-based nanofabrication and representative nanophotonic and nanoelectronic applications.** **(a)** Site specific assembly. The site-specific anchors on DNA origami templates enable spatial arrangement of nanoparticles/molecules or nanorods/nanotubes/nanowires with prescribed species, numbers, and orientations. **(b)** In-situ

synthesis. DNA origami templates mediate the adsorption/reaction of certain precursors on their surface or cavities. **(c)** Nanolithography. DNA origami structures as masks or stamps enable the transfer of their shapes or molecular patterns to other materials. **(d)** Low-loss plasmonic waveguide[54]. **(e)** Chiral plasmonic nanostructures[41]. **(f)** Reconfigurable 3D plasmonic metamolecules allowing dynamic circular dichroism changes[118]. **(g)** Single-molecule surface-enhanced Raman spectroscopic (SERS)[138]. **(h)** Nanoantennas for fluorescence enhancement[139]. **(i)** Engineering photonic crystal cavity (PCC) emission[136]. **(j)** DNA rotor driven by electric field. **(k-l)** Aligned carbon nanotubes[22] and tailored graphene patterns[23] enabling field effect transistor application.

**Nanophotonics and Nanoelectronics**

Many fascinating optical and electronic properties arise from the structural features with dimensions below the electromagnetic wavelengths (typically <100 nm). However, precise sculpturing of materials at that scale is challenging. DNA origami-templated nanostructures with high structural programmability at the nanometre level allow tailorable optical or electronic properties, including tuneable conductivity, plasmon coupling, Fano resonances, plasmonic chirality, thus holding great promise for applications in nanophotonics and nanoelectronics[21, 140].

Due to the localized surface plasmon resonance (LSPR), the photonic properties of a complex metal nanostructure composed of multiple nanoparticles are dependent on the particle size, shape, and interparticle spacing and configuration. Rigid DNA origami templates enable precise arrangement of heterogeneous plasmonic nanoparticles (e.g., coupling of AuNP and AgNP) of large sizes (e.g., >40 nm)[138], as well as small interparticle spaces (e.g., sub-5 nm) with little variability[141], leading to prominent yet predictable plasmon coupling and Fano resonances[138], which are thus suitable for studying nanoplasmonic effects. A DNA origami-templated multi-particle plasmon coupling chain showed ultrafast and low-dissipation energy transfer, which exemplifies a new route for the realization of plasmonic waveguides (Fig. 4d)[54, 142]. In addition, DNA origami-based nanofabrication enables complex asymmetric plasmonic nanoparticle assemblies[33, 41, 118, 143-149]), which interact differently with left and right circularly polarized light due to the structural chirality and the strong plasmon coupling, leading to pronounced circular dichroism in the visible spectrum (Fig. 4e,f).

By organizing multiple chromophores/fluorophores with distinct spectral properties, prescribed distances, orientations and/or donor-to-acceptor ratios, DNA origami platforms enable efficient light-harvesting antennas and photonic wires with long-range directional energy transfer[25, 26, 150, 151]. Precise positioning of Raman chromophores or fluorophores between plasmonic nanoparticles can generate quantitative surface-enhanced Raman spectroscopic (SERS) [138, 152] (Fig. 4g) or surface-enhanced fluorescence spectroscopic responses (Fig. 4h)[139, 153], which provide promising materials for single-molecule sensing when coupled to target-specific ligands and trigger-responsive dynamic DNA self-assemblies. Precise placement of fluorophore-labelled DNA origami onto lithographically patterned photonic crystal cavities allows engineering of their coupling and thereby digitally controllable cavity emission intensity(Fig. 4i)[136].

Given their intrinsic electric properties, DNA origami nanostructures can be manipulated by electric fields, which can thus serve as moving parts in electro-mechanical nanodevices (Fig. 4j)[56, 154, 155]. Moreover, DNA origami allows shaping and arrangement of diverse materials with different conductive/semi-conductive/dielectric properties, providing a new route to fabricate complex nanoelectronic modules and devices, such as nanowires with tuneable conductivity[130], field-effect transistors (FETs) based on spatially organized CNTs (Fig. 4k)[22, 156] or DNA-templated graphene nanoribbons (Fig. 4l)[23].

**Catalysis**

Biocatalytic transformations are central to the production of metabolites, biomolecules and energy conversion in living systems. In very early considerations about DNA nanotechnology, Seeman envisioned that DNA structures have the potential to organize proteins in spatially well-defined patterns, at first for structural analysis[1]. Later it has been realized that DNA nanostructures and in particular DNA origami offer an excellent platform for organizing enzymes spatially, due to the unique addressability of DNA origami[157].

The hypothesis motivating spatial organization of enzymes in cascades is that the diffusion of small molecule substrates between enzymes is proximity dependent and thereby also the rate of the cascade. The most common strategy for organizing enzymes (and proteins in general) in DNA nanostructures is to conjugate an oligonucleotide strand to each enzyme and in turn add the conjugates to an assembled DNA structure containing designed complementary single stranded domains[103, 158-164]. The hypothesis was however challenged by a study by Hess et al. in 2016, who argued that the negatively changed DNA structures, to which the enzymes are anchored, alter the pH in favour of the catalytic processes[165]. This argument, however, cannot account for all the proximity effect observed experimentally. There remains to be some uncertainty about the proximity effect and recent studies have pointed in different directions[160, 166].

In the examples given above the enzymes are confined in 1 or 2 dimensions, however another strategy is to confine enzymes in 3D, which has the potential to enhance both the channelling of intermediates and the impact of the origami structure on the enzyme. In some examples single proteins have been encapsulated in 3D origami structures[91, 167, 168], while a few cases have demonstrated encapsulation of enzyme couples. Fan and coworkers encapsulated the cascade in a DNA origami tube folded from a flat origami sheet[169], Kostainan and coworkers in an open-ended DNA honeycomb-lattice DNA origami tube[170], and Yan and coworkers encapsulated the same enzymes in a closed honeycomb lattice nanocage[110]. In all three cases significant rate enhancement of the enzyme cascade was observed as well as higher stability of the enzymes[171].

**Computation**

DNA is an information-carrying molecule with a high degree of thermodynamic and kinetic programmability, the latter owing to the rate tenability of toehold-mediated strand displacement reactions[14]. Due to these properties, DNA has been a popular substrate for *in vitro* signalling networks and molecular computation, starting with Adleman's implementation of the travelling salesman problem[172]. Rational sequence design – often the last step in multiple layers of abstraction[173]– finalizes the encoded algorithm and optimizes the emergent behaviour of DNA-based circuits. This has resulted in the embedding of intricate programs, such as Winfree's square-root calculator[174]. However, the scaling and computation speed of these dissipative circuits is limited by their lack of spatial organization[174]. Combined with the modularity of DNA-based interactions, it was therefore arguably inevitable for DNA-based computation to interface with the structural paradigm of DNA nanotechnology. DNA-origami structures provide a chassis for scaffolding, co-localizing and compartmentalizing circuit components[175]. Indeed, spatial organization of DNA-based circuits through the use of origami frameworks can be exploited to accelerate reaction rates[176], modulate stochiometry[175], restrict or promote specific pathways[177, 178] and outputs[37], compartmentalize distinct functional modules (e.g. sensing, computation, actuation)[179], decrease computation errors stemming from cross-talk, and increase sequence recyclability[176].

Although localization mostly presents optimization strategy for deterministic algorithms, it qualitatively alters the applicability of stochastic methods. For instance, Qian et al. designed a DNA-origami robot that sorts unordered cargo into distinct piles, the algorithm of which relies entirely on a random walk and localized targets[31]. Similarly, random walkers can be guided by their local landscape[104]. In another application, Chao et al. implemented a parallel depth-first search algorithm to solve mazes on DNA-origami sheets, based on randomly searching DNA navigators[32]. Alternatively, the results from DNA-based combinatorial selection can be localized on DNA origami structures, thereby coupling unique structural patterns to specific input signals[180]. Such applications would not be possible without spatial organization.

**Molecular Machines**

The structural stability of DNA origami relies mostly on Watson-Crick base pairing and base stacking, both of which are reversible interactions. Such reversibility has been exploited as a conscious design choice; for instance, a nanocontainer that opens and closes has more applications than one that only encapsulates[43, 91]. Over the past decade, there has been an influx of dynamic DNA origami devices that transition between two or more (semi-)stable states[181]. Dynamic DNA structures differentiate not just in the type of input trigger, but also in the number of accessible states, their actuation speed, and whether transitions are reversible[182]. A noteworthy example is the electrically-controlled rotatable arm that reversibly explores a continuum of states with only milliseconds of actuation time[56].

A major goal for DNA nanotechnology is to create molecular machinery and motors; systems that not just switch between states upon sensing some external change, but which are progressively fueled through a closed state-path and thereby generate change externally[183, 184]. This change can be in the form of potential energy (e.g. by establishing new chemical bonds) or kinetic energy (e.g. rupturing chemical bonds, translocation). In order to fabricate such machines, it will arguably be necessary to integrate multiple simpler devices, just like how different parts are combined to create macroscale machines[181]. A recent study by Salaita and colleagues combines catalytic walkers with steric direction; the resulting system converts potential energy from RNA-based fuel into unidirectional microscale movement of an origami roller[185]. Notably, this motor moves autonomously and without aid of any external gradient nor patterning.

**Drug Delivery**

DNA origami carriers have been demonstrated for the delivery of diverse therapeutic molecules and materials, including doxorubicin (Fig 5a)[186, 187], immunostimulatory nucleic acids[188], small interfering RNAs[189], etc., which can be readily loaded onto the carriers via diverse interactions. Moreover, DNA origami structures can serve as containers with docking sites in their interior or within dedicated cavities, protecting the payloads from the environment.

One outstanding challenge for drug delivery is to efficiently cross biological barriers to reach the drug targets with minimal off-target effects[190]. Previous studies have shown that well folded DNA nanostructures are more resistant to enzymatic degradation[191] than ss-/ds-DNAs and are capable of entering live mammalian cells through energy-dependent endocytic pathways like some viral particles[192-195]. The dependency of cellular uptake on size and shape has also been investigated owing to the customizability of DNA origami[196, 197]. At the animal level, DNA origami structures are found to passively accumulate in solid tumors due to the enhanced permeability and retention (EPR) effect, enabling tumor-targeting drug delivery[186]. In addition, DNA nanostructures can penetrate mouse or human skin with size-dependent efficacies, suggesting applications in transdermal drug delivery to melanoma tumor[198]. Recently, DNA origami structures were found to preferentially accumulate in the kidney of mouse, which shows promise to treat kidney injury (Fig. 5b)[57].

To endow the DNA nanostructures with active targeting ability, certain ligand molecules[199][190] with known receptors expressed on target cells have been incorporated with DNA origami carriers. Due to the addressability of DNA nanostructures, the species, numbers, density and orientation of these ligands can be precisely defined, allowing optimized cell targeting ability based on spatial pattern recognition by the cells[190].

Physiological and intracellular environments are highly heterogeneous and therefore call for smart carriers capable of sensing environmental stimuli at different delivery stages and switching their structures/properties adaptively. Recent progress in DNA origami-based nanomachines has shown promise in smart drug delivery. For example, the Church group developed a DNA origami-based nanorobot (Fig. 5c) with input logic based on two structure-switching aptamers[43]. Using AND logic, the nano-container can only be opened when both aptamers bind their corresponding targets. A later demonstration successfully applied this strategy in living cockroaches[200]. In another example, Li et al. constructed a DNA origami nanorobot that could be selectively unfolded by nucleolin enriched in tumor-associated blood vessels (Fig. 5d), which allowed local exposure of the encapsulated thrombins for intravascular thrombosis, resulting in tumor necrosis and inhibition of tumor growth in mice[29].

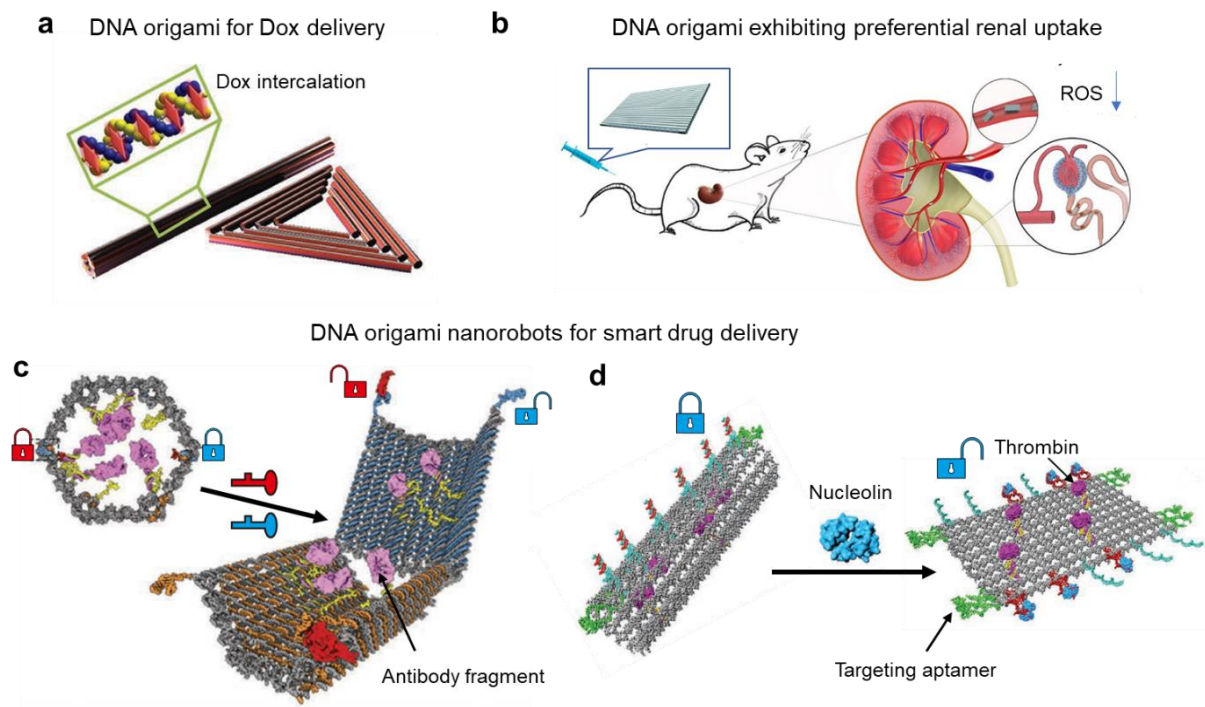

**Figure 5. Representative examples of DNA origami-based drug delivery applications. (a)** DNA origami for Dox delivery[186, 187]. **(b)** Preferential renal uptake of DNA origami enables the treatment of acute kidney injury[57]. **(c-d)**, DNA nanorobots for smart drug delivery[29, 43].

**Bioimaging and Biophysics**

DNA-origami structures serve as standards, markers, or structural support for the molecules of interest in biophysical studies. Placing designated number of fluorophores at predefined positions on DNA-origami structures (often sheets and rods) generates calibration standards for fluorescence microscopy to identity, count, and measure the spacing between molecular species. For example, Zhuang and Yin built a DNA-origami rotor with its centre attached to a dsDNA segment to track the dsDNA rotations induced by genome-processing enzymes (RecBCD and RNA polymerase)[58]. Similarly, stiff DNA rods enhance the signal-to-noise ratio of optical trap[51, 201] or FRET-based force measurement[202], leading to high-resolution study of weak (a few pico-Newton, pN) biological forces such as base stacking[51] and cell-substrate traction[202]. Labelling DNA filament with motor proteins and fluorophores enables the measurement of velocity, processivity and stall force of myosin[115, 203, 204], dynein and kinesin [205], both individually and in motor ensembles. In addition to optical imaging, DNA-origami structures have been exploited for AFM visualization of molecular motions[104, 106, 177, 206-211]. Clam-shaped DNA-origami structures have facilitated the EM analyses of nucleosome interaction and stability, using its open or closed conformations to signify various states of nucleosome assembly[83, 212-214]. Barrel or clamp shaped 3D DNA-origami structures have found applications in cryo-EM structural determination of proteins. Here, DNA nanostructures serve to define the thickness of vitrified ice sheet[215], create hydrophobic environment to stabilize membrane proteins[216], and orient DNA-binding proteins at desired rotation angles[217].

DNA origami creates artificial microenvironment to study molecular mechanisms of biological processes. An illustrative example is the study of multivalent antigen-antibody and protein-aptamer bindings on DNA-origami platform, where antigens and aptamers are organized in assorted arrangements to identify the molecular patterns that contribute to avidity[61, 63, 218, 219]. Crowding a dsDNA with protospacer-adjacent motifs has led to enhanced cas9/sgRNA binding and more efficient dsDNA cleavage[220]. Varying nucleoporin type and grafting density inside a DNA-origami cylinder have been shown to significantly impact the collective morphology and conductance of the intrinsically disordered proteins[221, 222]. Using DNA-origami-templated liposome formation techniques, liposomes and membrane proteins have been placed at defined distances and densities to study the biophysics of membrane dynamics[49, 223-225]. Flat or curved DNA origami surfaces outfitted with amphipathic molecules (e.g. cholesterol and peptides) can lead to membrane binding and deformation, useful for studying membrane mechanics[226-230].

## 5. Reproducibility and Data Deposition

The general standards of the DNA origami assembly have been continuously developed, ranging from the DNA origami designs, purification methods, to reconstruction models for TEM imaging, among others[82, 231]. To ensure a high reproducibility, several important aspects should be taken into account and carefully examined. First, a one-pot reaction can be used for the self-assembly of DNA origami. However, this often results in many by-products. A variety of the assembly conditions, such as the annealing procedure and the cationic strength can largely influence the yields of the target products and by-products. Optimization of the annealing procedure, especially the annealing temperature intervals, is crucial to achieving the target object with high yield. Furthermore, the cationic strength is another critical parameter for optimization to avoid DNA origami dissociation through electrostatic repulsion. The Tris-acetate-EDTA (TAE) buffer with $Mg^{2+}$ on the mM level is typically adopted in most protocols for the DNA origami assembly. Second, purification of the DNA origami is of great importance. There are mainly five purification methods, including PEG precipitation, gel purification, filter purification, ultracentrifugation, and size-exclusion chromatography. The most proper purification method for a particular experiment should be selected based on the yield and duration as well as the volume limitation, dilution, residuals, and damages[66, 82]. Finally, storage and stability of the DNA origami are also highly relevant. The storage temperature and cationic strength are both crucial for the stabilization of DNA origami. In general, DNA origami is thermally stable at temperatures of ~55°C or below in solution. It is worth mentioning that DNA origami was found to be stable over 85°C due to photo-crosslinking-assisted thermal stability[232]. It was also demonstrated that DNA origami could be lyophilized and stored under freezing conditions[233]. For the cationic strength, many approaches have been reported for protecting DNA origami from destabilization, especially at low $Mg^{2+}$ concentrations[234].

To ensure the data reproducibility, researchers are obliged to provide sufficient general information as well as detailed experimental conditions and procedures in publications. In addition, the experimental protocols and methods, assembly materials and sources, design and analysis software should also be carefully listed and described in detail. Data deposition in public repositories (such as Biorxiv) is highly recommended. Sharing the original data and detailed protocols in public are imperative and vital for the rigorous development of the field as well as will further push the advancement and broad interest of the DNA origami technique.

## 6. Limitations and Optimizations

The remarkable breadth of applications not only highlight the power of DNA origami, but also reveal roadblocks that need to be removed in order for the technology to reach its full potential. Somewhat surprisingly, the *first* limitation is the structural design, which, to this date, remains a major hurdle for those new to DNA nanotechnology and sometimes challenging even for DNA-origami experts. This is certainly not due to the lack of method development, as a long list of design and simulation tools[16, 46, 67, 68, 71-73, 75, 76, 78, 79, 82, 235-244] have been developed and made available to the public. Rather, the hurdle is caused by the fact that the more versatile design tools generally require a considerable amount of user input and technical know-hows, yet the better automated tools are typically geared towards specific types of construct. Ideally, we would enjoy a suite of software that streamlines the design and simulation process, serving both as a black box to convert simple geometrical models to DNA-origami designs and a sandbox for users to explore new design concepts. The field has made steady strides toward this goal by interfacing multiple design-simulation software, developing more user-friendly interfaces, and allowing for post-simulation touchup to optimize design iteratively. *Second,* a clear picture of the DNA-origami assembly mechanism remains elusive. Besides driving higher assembly yield of the target structures, the ability to clearly define DNA-origami folding pathways will enable rationally designed dynamic assemblies that can toggle between a few meta-stable conformations with low energy barriers in between - a feature found in many protein machineries. A number of high-quality studies have shed light on this long-standing question, including systematically testing DNA-origami design variants that lead to different folding outcomes[86, 98, 245], measuring the global thermal transition during DNA-origami assembly and disassembly[87], directly observing assembly intermediates[99], and monitoring the incorporation of selected staple strands[93, 94]. Many of these studies suggest a multi-stage, cooperative folding behavior. Future effort to depict such complicated mechanism will benefit from high-throughput analytical methods[246] and simulation framework[75, 76, 78, 79, 82, 237-239, 244] for DNA self-assembly. **Third**, the size of a DNA-origami structure is limited by the length of its scaffold strand, typically 7-8 kb long. To obtain larger structures, one has to use a longer scaffold[247, 248] and/or stitch multiple DNA-origami structures together[10, 11, 15, 249]. Both methods have been proven successful thanks to bacteriophage genome engineering and hierarchical DNA self-assembly[102, 250-252] via sticky-end cohesion or blunt-end base stacking, though the successes often come at the expense of assembly yield. Therefore, a related practical issue is how to generate staple strands in a cost-effective way to fold these massive structures (up to gigadalton scale) in large quantities. By far the most promising solutions are enzyme-mediated *in vitro* amplification of synthetic DNA oligonucleotides[253, 254] and biological production of DNA strands in bacteriophage[53, 253]. Using the latter approach, Dietz et al. produced several hundred milligrams of DNA-origami structures at a fraction of the cost of using conventional synthetic DNA[53]. **Fourth**, some intrinsic properties of DNA, such as its negative charge and susceptibility to enzyme degradation, may limit its applications, especially in biological environment. On the other hand, certain applications in physics and material science put the thermal and mechanical stability of DNA-origami structures to the test. When met with such challenges, a combination of DNA modification chemistry can come to rescue, for example photo-crosslinking DNA nucleotides[232, 255], wrapping exposed DNA surfaces with lipid bilayers[19], shielding DNA backbones with poly-cationic polymers[20, 256-260], and coating DNA with silica[18]. These modifications have helped DNA-origami structures survive low-salt, high-heat conditions, resist nuclease digestion, evade immune surveillance, prolong in vivo circulation, and avoid surface deformation. **Fifth**, although this is not a technical limitation, we cannot stress enough a technology's need for a healthy, vibrant ecosystem. The future of the DNA origami, and structural DNA nanotechnology at large, will be shaped by ambitious technology developers, who constantly push the technological frontiers only to shatter them, and a diverse user group, who brings new inspirations and challenges to fuel future innovations and keep the field relevant for the broad scientific community.

## 7. Outlook

**Molecular programming and automation.** DNA origami is a robust, sequence-programmable, nanometer-precise self-assembly technique, which is amenable to automation. It is not incidental that researchers in DNA nanotechnology and computing have dubbed their approach "molecular programming"[261]. Quite literally, DNA origami structures can be "programmed" using computer-aided tools[46, 67] without much knowledge about its chemical details - just as modern high-level computer programmers do not need to know what is going on at the hardware level. Arguably, the development of the caDNAno design program was one of the major catalysts for the field[67], which allowed even newcomers to design complicated supramolecular structures with a good chance of success.

It is therefore expected that one line of research will continue to be devoted to the "molecular programming" of nanostructures. This will involve an even stronger interconnection of computational design, automated synthesis and assembly of the structures, which could ultimately lead to a completely automated process for the generation of DNA-based nanostructures – potentially combined with on-demand DNA synthesis[262] (Fig. 6a).

**Chemistry for applications.** DNA origami has become popular because it meets the need for a molecular technology that enables the positioning of molecules and nanoparticles with nanometer-precision and with a defined stoichiometry. This allows to address a wide range of problems in nanoscience and in the life sciences, which are related to questions such as how to best arrange multiple interacting molecules or nanoparticles with respect to one another.

As the origami field moves more towards applications, most researchers will deal less with the refinement of the technique itself than with coping with the specifics of their application. Much future research will therefore be devoted to resolving chemical requirements of the applications such as making DNA origami chemically and physically stable, and developing molecular adaptors for functional components[263].

**Dynamic devices and robots.** One of the frontiers of DNA nanotechnology is the use of origami structures as components for dynamical "molecular robots". The complexity of origami structures allows integration of multiple functions in one device, which is required for molecular robots that involve the interplay of sensing, computing, and actuation modalities within a molecular system. One of the most promising applications for such systems will be in the context of biomedical nanorobotics, which will continue to be an intensive field of research over the next decade[43].

As parts of molecular assembly lines, origami-based nanorobots could perform "mechanical" synthesis of molecules by bringing components into proximity and appropriate orientation for reaction[264, 265], based on instructions read from a molecular tape. It is conceivable that catalytic origami structures could be developed that promote the reaction between two or more molecular components localized to the structures. Potentially, these reactions could be coupled to conformational changes of the origami structures, which could be either driven by some energy consuming chemical process, or physically using external fields[266 56 263].

**Multiscale molecular manufacturing.** Even though origami structures provide a means to control matter at the 10-100 nm length scale, it is unlikely that the technique will be applied for the creation of much larger structures. Instead, future research will be devoted to developing strategies for multiscale integration of DNA origami objects. This will involve the combination with "space-filling" materials other than DNA, and the self-assembly or printing of origami tectons into discrete or crystalline higher order structures using other interactions than DNA base-pairing[10] (Fig. 6b).

DNA origami will continue to be used to explore fundamental questions of self-assembly and self-organization, which may give rise to completely novel assembly strategies. While nature utilizes manifold non-equilibrium self-assembly processes, their use in the context of DNA nanostructures is still largely underexplored. Coupling of DNA self-assembly to energy-consuming processes will open up completely new possibilities for dynamically assembling[267], active molecular materials. One of the great visions in this context is the creation of molecular systems capable of self-replication[268].

**Generalizing the "origami idea".** The existing origami technique still has several shortcomings: for instance, DNA structures defined on lattices only have a limited spatial resolution. Another issue refers to the restricted chemical functionality of DNA, usually necessitating chemical modification of DNA staple strands. Finally, for some applications it may be desirable to produce origami-like structures inside living cells.

Future efforts will be devoted to the question whether the "origami idea" can be extended in scope, applied to other molecules, and made to work in vivo. Several groups have already started to create "lattice-free" origami structures[16] that allow folding of structures in wireframes or along arbitrary object-filling curves. In order to increase chemical versatility, it will be interesting to explore whether DNA origami concepts can be applied to the design of protein nanostructures. Hybrid approaches are conceivable, in which DNA or RNA scaffolds support the association of proteins into heteromultimeric complexes. An exciting alternative is the extension of the chemical capabilities of origami using xeno-nucleic acids (XNAs) with different backbones and unnatural base-pairs[269, 270].

**In vivo production.** While the production of origami scaffolds and staples inside cells has been successfully demonstrated[53], it is hard to imagine that annealing of hundreds of oligonucleotides would work in this context. Intramolecular folding of structures here appears to be more promising. Along these lines, there has been great progress in the development of single-stranded DNA origami[12], and also RNA origami[271], which is based on co-transcriptional folding of RNA.

Single-stranded RNA or DNA origami can be produced by template-directed enzymatic synthesis. As XNAs are designed to be compatible with biopolymerization processes, they could also be included in enzymatically produced nanostructures. An exciting perspective could thus be an expansion of the genetic alphabet to code for XNA-based nanostructures that can be synthesized in vivo. First experiments have shown that bacteria containing unnatural base-pairs can replicate, and thus bacterial strains could be developed that act as XNA origami producers[272]. Coupling the production of (DNA, RNA or XNA) origami to reproduction finally opens up the possibility for evolution, which could result in even more complex structures with completely novel functionalities (Fig. 6c).

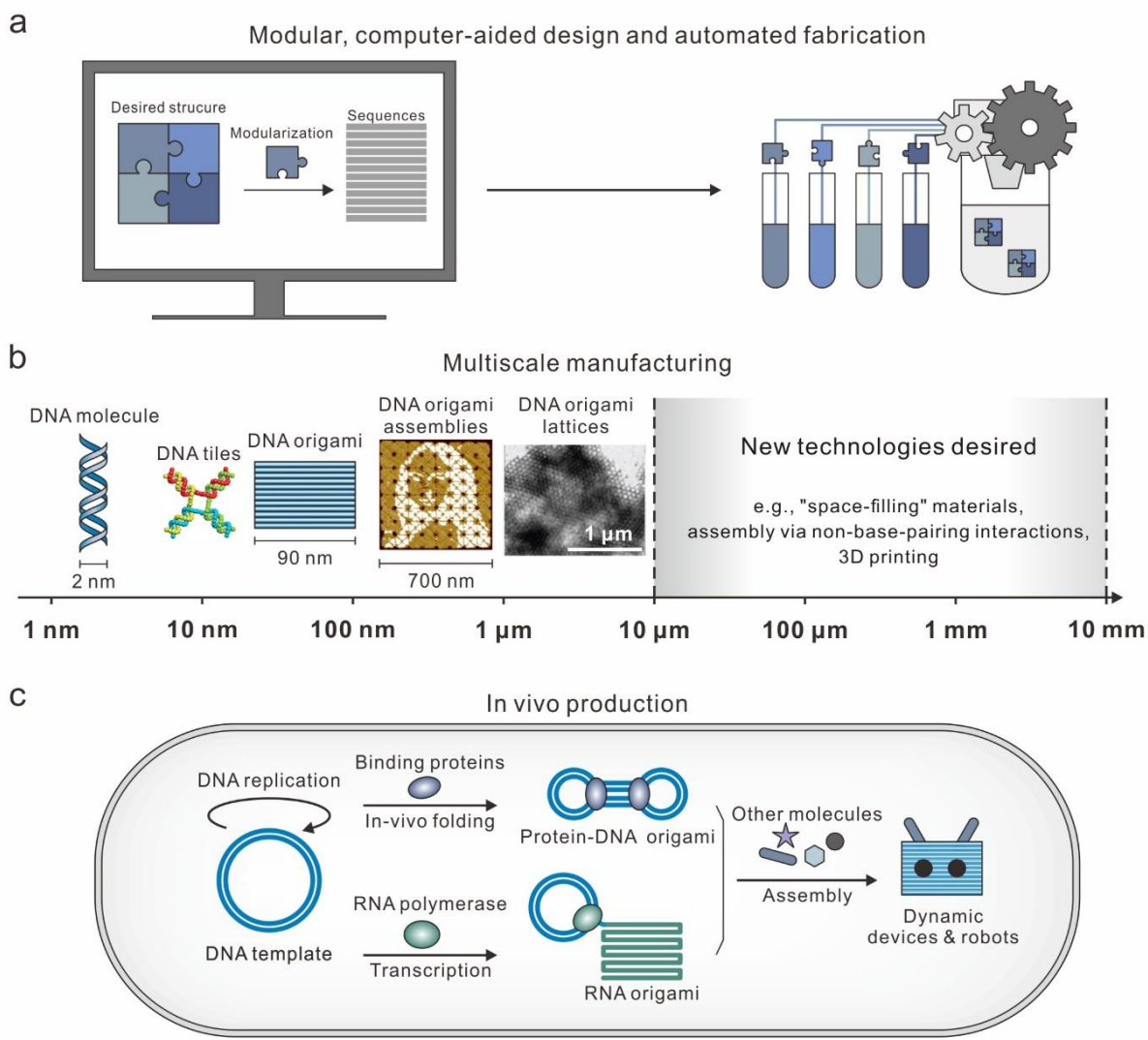

**Figure 6. Outlook of DNA origami technology.** **(a)** Modular, computer-aided design, and automated fabrication. **(b)** Multiscale manufacturing. **(c)** In-vivo production of DNA/RNA origami and further assembly of dynamic devices and robots.

# Acknowledgement

Parts of this work were supported by the National Natural Science Foundation of China (21991134, 21834007) and the Shanghai Municipal Science and Technology Commission (19JC1410300).